\documentclass[12pt]{article}
\usepackage{amssymb,amsmath,epsfig}
\begin{document}
\parindent=0pt
\parskip=6pt
\rm

\vspace*{0.5cm}

\begin{center}

\normalsize {\bf HOMOGENEOUS PHASE OF COEXISTENCE OF SPIN-TRIPLET
SUPERCONDUCTIVITY AND FERROMAGNETISM}

D. V. SHOPOVA$^{\ast}$ and D. I. UZUNOV

{\em CPCM Laboratory, G. Nadjakov Institute of Solid State Physics,\\
Bulgarian Academy of Sciences, BG-1784 Sofia, Bulgaria.} \\
\end{center}

$^{\ast}$ Corresponding author: sho@issp.bas.bg

\vspace{0.5cm}

order parameter profile.


\vspace{0.3cm}

\normalsize
\begin{abstract}

The coexistence of a homogeneous (Meissner-like) phase of spin-triplet
superconductivity and ferromagnetism is investigated within the
framework of a phenomenological model of spin-triplet ferromagnetic
superconductors. The results are discussed in view of application to
metallic ferromagnets as UGe$_2$, ZrZn$_2$,  URhGe, and Fe.
\end{abstract}

{\bf 1. Introduction}

Experiments at low temperatures and high pressure have indicated a
coexistence of ferromagnetism and superconductivity in the metallic
compounds UGe$_2$~\cite{Saxena:2000,Huxley:2001,Tateiwa:2001},
ZrZn$_2$~\cite{Pfleiderer:2001}, URhGe~\cite{Aoki:2001} and also in
Fe~\cite{Shimizu:2001}. In contrast to other superconducting materials
(see, e.g., Refs.~\cite{Blount:1979, Varma:1997}), in these metals the
phase transition temperature to the ferromagnetic state is higher than
the phase transition temperature to the superconducting state and  the
superconductivity  not only coexists with the ferromagnetic order but
is enhanced by it.  It is widely accepted~\cite{Saxena:2000,
Saxena:2001} that this superconductivity can be most naturally
understood as a spin-triplet rather than a spin-singlet pairing
phenomenon (see, e.g., Ref.~\cite{Mineev:1999}). The experiments
~\cite{Shimizu:2001} on high-pressure crystal modification of Fe, which
has a hexagonal closed-packed structure, are also interpreted
~\cite{Saxena:2000}  in favour of the appearance of same unconventional
superconductivity.  Note, that both vortex and Meissner
superconductivity phases~\cite{Shimizu:2001} are found in the
high-pressure crystal modification of Fe where the strong
ferromagnetism of the usual bcc iron crystal probably
disappears~\cite{Saxena:2001}.

The  phenomenological theory that explains coexistence of
ferromagnetism and unconventional spin-triplet superconductivity of
Landau-Ginzburg type was derived~\cite{Machida:2001, Walker:2002} on
the basis of general symmetry  group arguments . It describes the
possible low-order coupling between the superconducting and
ferromagnetic order parameters and establishes several important
features of the superconducting vortex state in the ferromagnetic phase
of unconventional ferromagnetic superconductors ~\cite{Machida:2001,
Walker:2002}. Both experimental and theoretical problems of the
ferromagnetism in spin-triplet superconductors seem to be quite
different from those in conventional (s-wave) ferromagnetic
superconductors~\cite{Blount:1979, Varma:1997}.

In this letter we investigate the conditions for the occurrence of the
homogeneous (Meissner-like) phase of coexistence of spin-triplet
superconductivity and ferromagnetism. Such a phase of coexistence may
occur at relatively small magnetization and at zero external magnetic
field. Taking in mind this circumstance and using model considered in
Refs.~\cite{Machida:2001, Walker:2002} we show that the phase
transition to the superconducting state in ferromagnetic
superconductors can be either of first or second order and this depends
on the model parameters that correspond to the particular substance.
Our investigation is based on the mean-field
approximation~\cite{Uzunov:1993} as well as on familiar results about
the possible phases in nonmagnetic superconductors with triplet
($p$-wave) Cooper pairs~\cite{Blagoeva:1990, Uzunov:1990}. We neglect
all anisotropies, usually given by the respective additional Landau
invariants and gradient terms~\cite{Mineev:1999} in the Ginzburg-Landau
free energy of unconventional superconductors. The reasons is that the
inclusion of crystal anisotropy is related with lengthy formulae and a
multivariant analysis which will obscure our main aims and results. Let
us emphasize that the present results should be valid in the same or
modified form when the crystal anisotropy is properly taken into
account. We have to mention also that there is a formal similarity
between the phase diagram obtained in our investigation and the phase
diagram of certain improper ferroelectrics~\cite{Gufan:1980}.

{2. \bf Model}

We consider the Ginzburg-Landau free energy~\cite{Machida:2001,
Walker:2002} $F=\int d^3 x f(\psi, \vec{{\cal{M}}})$, where
\begin{equation}
\label{eq1} f = \frac{\hbar^2}{4m} (D^{\ast}_j\psi)(D_j\psi) +
a_s|\psi|^2 + \frac{b}{2}|\psi|^4 + a_f\vec{{\cal{M}}}^2 +
\frac{\beta}{2}{\cal{M}}^4 + i\gamma_0 \vec{{\cal{M}}}.(\psi\times
\psi^*)\;.
\end{equation}
In Eq.~(\ref{eq1}), $D_j =(\nabla - 2ieA_j/\hbar c)$, and $A_j$ ($j =
1,2,3$) are the components of the vector potential $\vec{A}$ related
with the magnetic induction $\vec{B} = \nabla \times \vec{A}$. The
complex vector $\psi = (\psi_1,\psi_2,\psi_3) \equiv \{\psi_j\}$ is the
superconducting order parameter, corresponding to the spin-triplet
Cooper pairing and $\vec{{\cal{M}}}= \{{\cal{M}}_j\}$ is the
magnetization. The coupling constant $\gamma_0 = 4\pi J>0$ is given by
the ferromagnetic exchange parameter $(J>0)$. Coefficients $a_s =
\alpha_s(T-T_s)$ and $a_f = \alpha_f(T-T_f)$ are expressed by the
positive parameters $\alpha_s$ and $\alpha_f$ as well as by the
superconducting $(T_s)$ and ferromagnetic $(T_f)$ critical temperatures
in the decoupled case, when ${\cal{M}}\psi_i\psi_j$-interaction is
ignored; $b > 0$ and $\beta > 0$ as usual.

We assume that the magnetization ${\cal{M}}$ is uniform, which is a
reliable assumption outside a quite close vicinity of the magnetic
phase transition but keep the spatial ($\vec{x}-$) dependence of
$\psi$. The reason is that the relevant dependence of $\psi$ on
$\vec{x}$ is generated by the diamagnetic effects arising from the
presence of ${\cal{M}}$ and the external magnetic field
$\vec{H}$~\cite{Machida:2001,Walker:2002} rather than from fluctuations
of $ \psi $ (this effect is extremely small and can be safely ignored).
First term in~(\ref{eq1}) will be still present even for $\vec{H} = 0$
because of the diamagnetic effect created by the magnetization
$\vec{{\cal{M}}} = \vec{B}/4\pi > 0$. As we shall investigate the
conditions for the occurrence of the Meissner phase where $\psi$ is
uniform, the spatial dependence of $\psi$ and, hence, the first term in
r.h.s. of~(\ref{eq1}) will be neglected. This approximation should be
valid when the lower critical field $H_{c1}(T)$ is greater than the
equilibrium value of the magnetization ${\cal{M}}$ in the phase of
coexistence of superconductivity and ferromagnetism.

One may take advantage of the symmetry of  model~(\ref{eq1}) and avoid
the consideration of equivalent thermodynamic states that occur as a
result of the respective continuous symmetry breaking at the phase
transition point but have no effect on thermodynamics of the system. We
shall assume that the magnetization vector is along the  $z$-axis:
$\vec{{\cal{M}}} = (0,0,{\cal{M}})$, where ${\cal{M}}\geq 0$. We find
convenient to use the following notations: $\varphi_j =b^{1/4}\psi_j$,
$\varphi_j = \phi_j\mbox{exp}(\theta_j)$, $M = \beta^{1/4}{\cal{M}}$,
$\gamma= \gamma_0/ (b^2\beta)^{1/4}$, $r = a_s/\sqrt{b}$, $t =
a_f/\sqrt{\beta}$, $\phi^2 = (\phi_1^2 + \phi_2^2 + \phi_3^2)$, and
$\theta = (\theta_2-\theta_1)$.

We shall not dwell on the metastable and unstable phases described by
the model~(\ref{eq1})~\cite{Shopova:2003} as well as on the vortex
phase~\cite{Machida:2001,Walker:2002} corresponding to $|\vec{B}| >
H_{c1}$. Rather we consider the stable homogeneous phases at zero
external magnetic field ($\vec{H} = 0$) that are described by uniform
order parameters ${\cal{M}}$ and $\psi$. We shall essentially use the
condition $T_f > T_s$.

{\bf 3. Results and discussion}

The model~(\ref{eq1}) describes three stable homogeneous phases: (1)
normal phase ($\phi_j = M = 0$) (hereafter referred to as N-phase), (2)
ferromagnetic (FM-) phase ($\phi_j = 0$, $M^2 = -t > 0$), and (3) a
phase of coexistence of superconductivity and ferromagnetism
(FS-phase), given by $\phi_3 = 0$, $\theta = 2\pi(k - \ 1/4)$, $(k = 0,
\pm 1,...)$, $\phi^2 = (-r + \gamma M) > 0$, $6M^2>(\gamma^2-2t)$, and
\begin{equation}
\label{eq2} \frac{\gamma r}{2} = \left(\frac{\gamma^2}{2}-t \right)M -
M^3\;.
\end{equation}
It is not difficult to determine the domains of existence and stability
of the phases N, FM, and FS.  Note, that here we use the term
``condition of stability" to indicate the necessary condition of
stability when the respective phase corresponds to a minimum of the
free energy, i.e., in both cases of stable and metastable states. When
a phase corresponds to a global minimum of the free energy (a
sufficient condition of stability) it will called a ``stable phase."
Thus we easily find the following existence and stability regions: ($t
> 0$, $r> 0)$ for the N-phase, ($t < 0$, $r > \gamma M$) - for FM. In
order to obtain the same domain for FS one should consider
Eq.~(\ref{eq2}) together with the additional existence and stability
conditions corresponding to this phase: $\gamma M > r$ and $3M^2 >
M_0^2$, where $M_0 > 0$ is defined by the $r(M) = 0$; see
Eq.~(\ref{eq2}).

  Eq.~(\ref{eq2}) is shown in Fig.~1 for $\gamma =
1.2$ and $t = -0.2$. For any $\gamma > 0$ and $t$, the stable FS
thermodynamic states are given by $r (M) < r_m = r(M_m)$ for $M > M_m >
0$, where $M_m$ corresponds to the maximum of the function $r(M)$.
Functions $M_m(t)$ and $M_0(t) = (-t + \gamma^2/2)^{1/2} =
\sqrt{3}M_m(t)$ are drawn in Fig.~2 for $\gamma = 1.2$.  Functions
$r_m(t) = 4M_m^3(t)/\gamma$ for $t < \gamma^2/2$ (the line of circles
in Fig.~3) and $r_e(t) = \gamma|t|^{1/2}$ for $t < 0$ (the dotted in
Fig.~3) define the borderlines of existence and stability of FS. %
\begin{figure}
\begin{center}
\epsfig{file=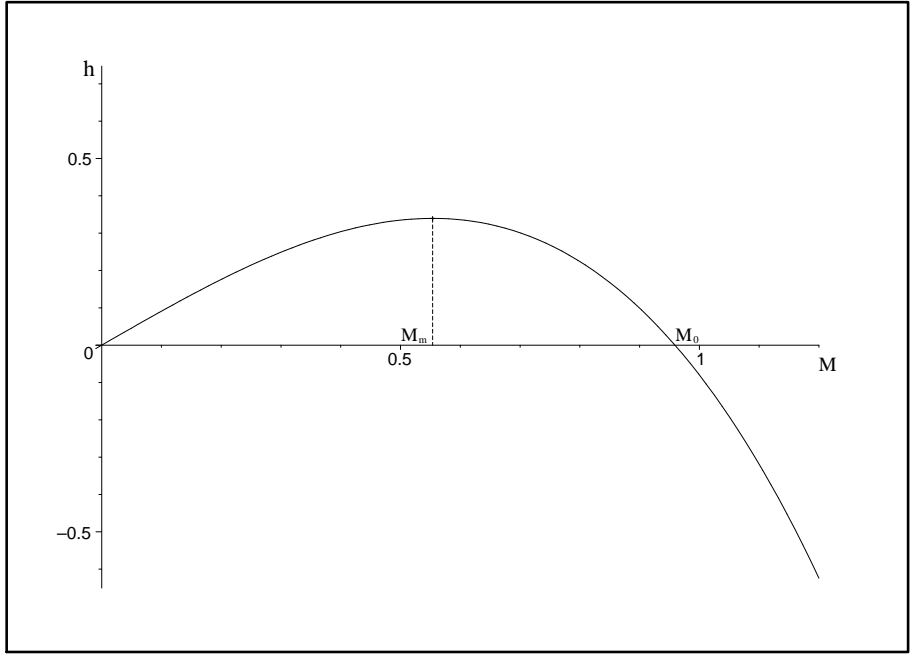,angle=-90, width=12cm}
\end{center}
\caption{$h=\gamma r/2$ as a function of $M$ for $\gamma = 1.2$, and $t
= -0.2$.} \label{su203f1.fig}
\end{figure}
\begin{figure}
\begin{center}
\epsfig{file=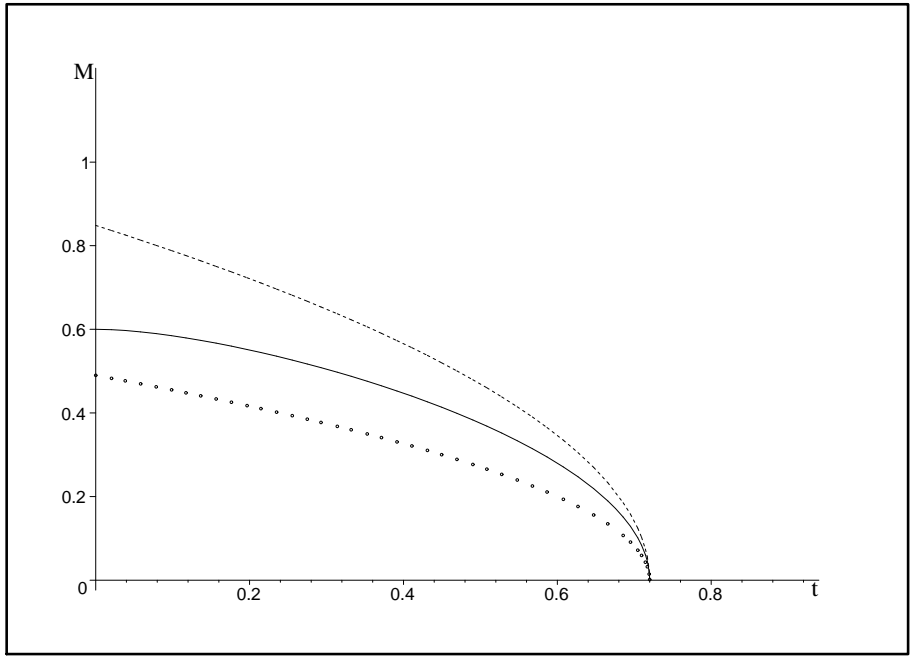,angle=-90, width=12cm}
\end{center}
\caption{$M$ versus $t$ for $\gamma = 1.2$: the dashed line represents
$M_0$, the solid line represents $M_{\mbox{eq}}$, and the dotted line
corresponds to $M_m$.} \label{su203f2.fig}
\end{figure}

In the region on the left of the point B (Fig.~3) with coordinates
($-\gamma^2/4$,$\gamma^2/2$), FS satisfies the existence condition
$\gamma M > r$ only below the dotted line $\left[r < r_{e}\right]$. In
the domain confined between the lines of circles and the dotted line on
the left of the point B the stability condition for FS is satisfied but
the existence condition is broken. The inequality $r \geq r_e(t)$ is
the stability condition of FM for $ 0 \leq (-t) \leq \gamma^2/4$. For
$(-t) > \gamma^2/4$ the FM phase is stable for all $r \geq r_e(t)$. The
dotted line on the left of the point B, i.e. for $(-t) > \gamma^2/4$),
is a line of the second order FM-FS phase transition. On this line the
equilibrium order parameters are given by $\phi_j = 0$ and $M_{eq} =
\sqrt{|t|}$. Therefore, the phase diagram for $(-t) > \gamma^2/4$ is
clarified for any $r$. When $r < 0$ the FS phase is stable and is
described by the function $r(M)$ for $M > M_0$, as shown  in Fig.~3.

\begin{figure}
\begin{center}
\epsfig{file=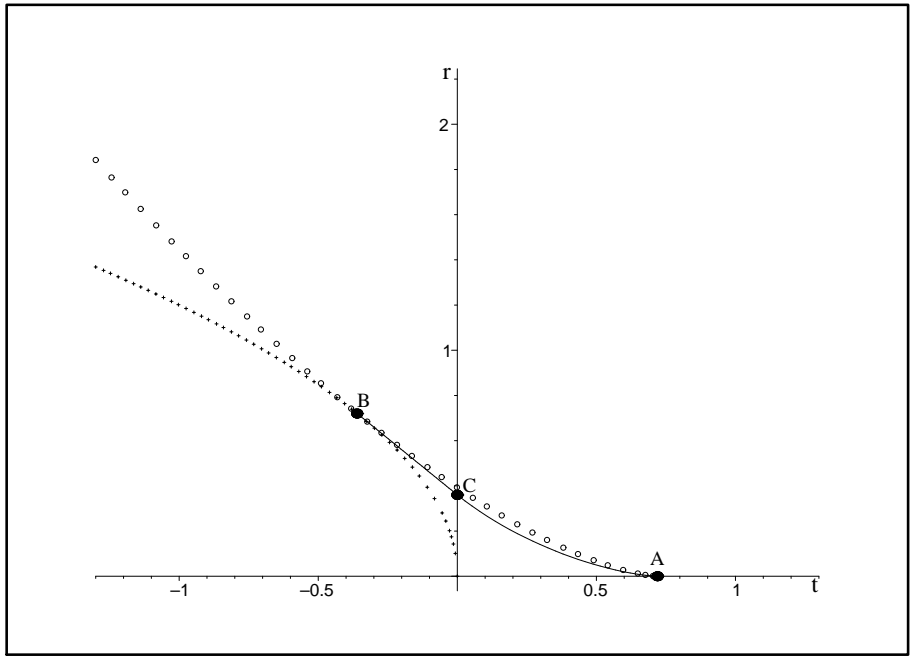,angle=-90, width=12cm}\\
\end{center}
\caption{The phase diagram in the plane ($t$, $r$) with two tricritical
points (A and B) and a triple point $C$; $\gamma = 1.2$.}
\label{su203f3.fig}
\end{figure}
The part of the $t$-axis given by $r=0$ and $t > \gamma^2/2$ in Fig.~3
is a phase transition line of second order that describes the N-FS
transition. The same transition for $0 < t < \gamma^2/2$ is represented
by the solid line AC which is the equilibrium transition line of a
first order phase transition. This equilibrium transition curve is
given by the function
\begin{equation}
\label{eq3} r_{eq}(t) = \frac{1}{4}\left[3\gamma - \left(\gamma^2 + 16t
\right))^{1/2}\right]M_{eq}(t),
\end{equation}
where
\begin{equation}
\label{eq4} M_{eq}(t) = \frac{1}{2\sqrt{2}}\left[\gamma^2 - 8t +
\gamma\left(\gamma^2 + 16t \right)^{1/2}\right]^{1/2}
\end{equation}
is the equilibrium value (jump) of the magnetization. The order of the
N-FS transition changes at the tricritical point A.

The domain above the solid line AC and below the line of circles for $
t > 0$ is the region of a possible overheating of FS. The domain of
overcooling of the N-phase is confined by the solid line AC and the
axes ($t > 0$, $r >0$). At the triple point C with coordinates [0,
$r_{eq}(0) = \gamma^2/4$] the phases N, FM, and FS coexist. For $t < 0$
the straight line
\begin{equation}
\label{eq5} r_{eq}^* (t) =  \frac{\gamma^2}{4} + |t|,\;\;\;\;\;\;
-\gamma^2/4 < t < 0,
\end{equation}
describes the extension of the equilibrium phase transition line of the
N-FS first order transition to negative values of $t$. For $t <
(-\gamma^2/4)$ the equilibrium phase transition FM-FS is of second
order and is given by the dotted line on the left of the point B (the
second tricritical point in this phase diagram). Along the first order
transition line $r_{eq}^{\ast}(t)$, given by~(\ref{eq5}), the
equilibrium value of $M$ is $M_{eq} =\gamma/2$,  which implies an
equilibrium order parameter jump at the FM-FS transition equal to
($\gamma/2 - \sqrt{|t|}$). On the dotted line of the second order FM-FS
transition the equilibrium value of $M$ is equal to that of the FM
phase ($M_{eq} = \sqrt{|t|}$). At the triple point C the phases N, FM,
and FS coexist.

In conclusion, we have demonstrated that the model (1) of ferromagnetic
spin-triplet superconductors gives a quite complex phase diagram
containing three ordered phases, two types of phase transitions, and
two tricritical points, and a triple point. Further considerations of
the effect of additional terms in the free energy~(\ref{eq1}) such as
terms of the type $\vec{\cal{M}}^2|\psi|^2$ and/or terms describing the
Cooper pair and crystal anisotropy~\cite{Mineev:1999, Blagoeva:1990}
may give more information about the shape of the phase diagram outlined
in the present paper.

{\bf Acknowledgments:} DIU thanks the hospitality of MPI-PKS-Dresden.
Financial support through {\em Scenet} (Parma) and collaborative
project with JINR-Dubna is also acknowledged.


\newpage
\begin{thebibliography}{ll}
\bibitem{Saxena:2000}
S. S. Saxena, P. Agarwal, K. Ahilan, F. M. Grosche, R. K. W.
Haselwimmer, M.J. Steiner, E. Pugh, I. R. Walker, S.R. Julian, P.
Monthoux, G. G. Lonzarich, A. Huxley. I. Sheikin, D. Braithwaite, and
J. Flouquet,  {\em Nature} (London) {\bf 406}, 587 (2000).
\bibitem{Huxley:2001}
A. Huxley, I. Sheikin, E. Ressouche, N. Kernavanois, D. Braithwaite, R.
Calemczuk, and J. Flouquet, {\em Phys. Rev.} {\bf B 63}, 144519-1
(2001).
\bibitem{Tateiwa:2001}
N. Tateiwa, T. C. Kobayashi, K. Hanazono, A. Amaya, Y. Haga. R. Settai,
and Y. Onuki, {\em J. Phys. Condensed Matter} {\bf 13}, L17 (2001).

\bibitem{Pfleiderer:2001}
C. Pfleiderer, M. Uhlatz, S. M. Hayden, R. Vollmer, H. v. L\"ohneysen,
N. R. Berhoeft, and G. G. Lonzarich, {\em Nature} (London) {\bf 412},
58 (2001).
\bibitem{Aoki:2001}
D. Aoki, A. Huxley, E. Ressouche, D. Braithwaite, J. Flouquet, J-P..
Brison, E. Lhotel, and C. Paulsen,{\em Nature} (London) {\bf 413}, 613
(2001).
\bibitem{Shimizu:2001}
K. Shimizu, T. Kimura, S. Furomoto, K. Takeda, K. Kontani, Y. Onuki and
K. Amaya,{\em Nature} (London) {\bf 412}, 316 (2001).
\bibitem{Blount:1979}
E. I. Blount and C. M. Varma, {\em Phys. Rev. Lett.} {\bf 42}, 1079
(1979).
\bibitem{Varma:1997}
T. K. Ng and C. M. Varma, {\em Phys. Rev. Lett.} {\bf 78}, 330 (1997).
\bibitem{Mineev:1999}
V. P. Mineev, K. V. Samokhin, {\em Introduction to Unconventional
Superconductivity} (Gordon and Breach, Amsterdam, 1999).
\bibitem{Saxena:2001}
S. S. Saxena and P. B. Littlewood, {\em Nature} (London) {\bf 412}, 290
(2001).
\bibitem{Machida:2001}
K. Machida and T. Ohmi, {\em Phys. Rev. Lett.} {\bf 86}, 850 (2001).
\bibitem{Walker:2002}
M. B. Walker and K. V. Samokhin, {\em Phys. Rev. Lett.} {\bf 88},
204001-1 (2002).
\bibitem{Uzunov:1993}
D. I. Uzunov, {\em Theory of Critical Phenomena} (World Scientific,
Singapore, 1993).
\bibitem{Blagoeva:1990}
E. J. Blagoeva, G. Busiello, L. De Cesare, Y. T. Millev, I. Rabuffo,
and D. I. Uzunov, {\em Phys. Rev.} {\bf B 40}, 7321 (1990).
\bibitem{Uzunov:1990}
D. I. Uzunov, in: {\em Advances in Theoretical Physics}, ed. by E.
Caianiello (World Scientific, Singapore, 1990) p. 96.
\bibitem{Gufan:1980}
Yu. M. Gufan and V. I. Torgashev, {\em  Sov. Phys. Solid State} {\bf
22}, 951 (1980)  [{\em Fiz. Tv. Tela} (Leningrad) {\bf 22}, 1629
(1980)].
\bibitem{Shopova:2003}
D. V. Shopova and D. I. Uzunov, {\em Phys. Lett. A} (2003) in press.
\end{thebibliography}
\end{document}